\documentclass[1p]{elsarticle}
\usepackage{graphicx,t1enc}
\usepackage{amsmath}
\usepackage{nicefrac}
\usepackage[colorlinks=true,linkcolor=blue,urlcolor=blue,]{hyperref}

\biboptions{numbers,sort&compress}

\journal{Physica A}

\begin{document}
\begin{frontmatter} 
\title{An energy-based natural selection model}

\author[cab,ib]{Noam Abadi}
\ead{noam.abadi@ib.edu.ar}
\author[cab,ib,conicet]{Guillermo Abramson}
\ead{abramson@cab.cnea.gov.ar}

\address[cab]{Centro At\'omico Bariloche (CNEA), R8402AGP Bariloche,  
Argentina}
\address[ib]{Instituto Balseiro, Universidad Nacional de Cuyo, Argentina}
\address[conicet]{Consejo Nacional de Investigaciones Cient\'{\i}ficas y T\'ecnicas, Argentina}

\begin{abstract}
Genetic information and environmental factors determine the path of an individuals life and therefore, the evolution of its entire species. We have succeeded in proposing and studying a model that captures this idea. In our model, a renewable resource extended throughout the environment provides the energy necessary to sustain life, including movement and reproduction. Since the resource doesn't regrow immediately, it generates competition between individuals and therefore provides a natural selection pressure from which evolution of the genetic traits is observed. As a result of this, several phenomena characteristic of living systems emerge from this model without having to introduce them explicitly. These include speciation and punctuated equilibrium, competitive exclusion, and altruistic behaviour from selfish rules.

\end{abstract}

\date{\today}


\end{frontmatter}


\section{Introduction}
\label{sec:int}

Darwin's theory of evolution~\cite{Darwin} was a huge step towards answering questions regarding the origins of life and providing a unifying framework for its diversity as observed on Earth. However, the theory is so broad in its reach that many aspects of it are still under study. For example, it has become increasingly accepted that the evolution of the biosphere needs an integrated description of both the biological agents as well as their resources~\cite{smith2016}. Indeed, the main agents and processes responsible for the evolution, namely speciation and extinction, are the result of interactions, and interactions are often mediated by shared resources.

The main area where modern understanding transcends Darwin's is genetics. We understand many details of the inheritance of traits, from the molecular to the physiological levels. Yet, the relation between genetically inherited traits and the probability of producing offspring in a certain environment is not completely clear. Similarly, it is not obvious what general features of extinction, survival, and eventually, speciation should arise from the complex interaction of the many mechanisms involved. Many of these problems have been studied in recent decades from a theoretical point of view, with a variety of tools rooted in the theory of complex systems \cite{Kauffman,bak1993,abramson1998,elyacoubi2006}. 

Moreover, living systems are chemical, and their dynamics is driven by free energy.	At the core of life, there seems to be just a chemical metabolism, which nowadays is a complex network of reaction pathways, involving both equilibrium and out-of-equilibrium processes that drive the flow of free energy through the system. Indeed, there is no indication that individuality or replication played a role in early chemical (pre-biotic) organization \cite{oparin,prado2018}. Even if they perhaps do not \emph{define} life, individuality, replication, and selection arose at some moment in the evolution of living systems as persistent and universal distinctions. In this work, we propose a model in which individual agents have traits, determined by genes, that determine the balance between energy input and output. We study how a population of such creatures evolves, in a scenario of competition for a spatially extended common resource that provides the necessary energy. The capacity to move is one such trait, and it allows creatures to change their position when they exhaust their local resource, albeit at a cost in energy. Likewise, body mass is energetically expensive to keep alive and also to carry around but guarantees a reserve in case of resource scarcity. Since both traits have benefits and drawbacks, we expect them to have optimal values, which the system should achieve in a stationary state, in a context of competition mediated by the common resource. 

The model is similar to latent energy environment (LEE) models~\cite{menczer1996a,menczer1996b}, but with two key differences: first, the resource is entirely equivalent to a creatures energy instead of components that need to be combined with others to produce it. This allows the second distinction, which is that individuals have no neural network to act as a brain and allow them to learn throughout their lives. The advantage of this is that the model can address questions regarding primordial life forms.

As opposed to artificial selection, which evolves towards the \textit{optimization} of a certain function~\cite{GeneticAlgorithms}, natural selection evolves under a certain \textit{constraint}, which we propose is the limitation of energy. Therefore, the model does not prescribe any landscape of fitness, nor any function to be optimized. Traits are inherited from one generation to the next with small, random mutations, that may change their performance and the corresponding energy balance. This makes the population undergo a natural selection process as the agents interact through the common resource. Selection is not modeled, but an emergent mechanism that provides a certain organization to the ecosystem. 

We show that phenomena commonly associated with natural selection appear to be intrinsically tied to it in our model. The first one is speciation, in the form of two possible sets of traits in a stationary state, which is achieved by mechanisms akin to those of punctuated equilibrium. The second one is that the main behavior throughout the population is much more altruistic than what is expected from individual rules, as suggested by the selfish gene theory, both in the amount of resource consumed and life expectancy. We also provide a mean-field description of the resource dynamics, which shows important consequences of its spatial extension and limited growth.

\section{The hydra model}
\label{sec:age_hyd}

We will proceed to formulate our model\footnote{We call it the ``hydra model'' because of superficial similarity with the freshwater genus of the same name: tumbling motion, asexual reproduction by budding and potential immortality. Details of the model, either particular parameter values or results, do not pretend to correspond with actual specific features of the life history of \emph{Hydra sp.}} based on the following key ingredients of an evolutionary system as identified by Kaneko~\cite{kaneko2006}: 1) Genotype, phenotype, and environment; 2) Fitness depends on the phenotype and the environment; 3) Only genes are transferred to the offspring; 4) Mutation of genes occurs randomly without any specific direction. Moreover, in the formulation that we describe below, fitness is not a prescribed function but an emergent feature of the dynamics, as will be discussed further on.

Consider a grid of size $L \times L$, and a resource $f$ extended over it. Let us suppose that, in the absence of consumers, its dynamics obeys a locally logistic equation~\cite{log_eq}:
	\begin{equation}
		\frac{d\, f(\mathbf{x})}{dt} = r f(\mathbf{x}) \left( 1 - \frac{f(\mathbf{x})}{K} \right).
		\label{eq:res_dyn}
	\end{equation}
A population of individuals lives on the grid and moves around it, feeding off the resource to obtain the energy $E$ needed to stay alive and storing it in their bodies. For each creature this energy evolves according to intake and consumption
	\begin{equation}
		\frac{dE(t)}{dt} = \varphi(t) - C(t),
		\label{eq:phi-cons}
	\end{equation}
where the feeding function $\varphi$ is the amount of resource that the creature gets from the cell on which it is located, and the consumption $C$ is the amount of energy it uses to fulfill its bodily needs. Equation (\ref{eq:phi-cons}) states that energy cannot be created inside a creature's body, and therefore has little meaning without modelling $\varphi$ and $C$. To this end, we define a genotype ${\bf G} = \{ M, P, s \}$, which consists of the adult body size, $M$, the adult probability of moving to a nearest neighbour cell, $P$, and the growth rate from birth to adulthood, $s$. Upon reproduction, the genotype is transmitted to the next generation with small mutations, allowing the population to evolve. The genotype is expressed as a time (age) dependent phenotype ${\bf F}(t) = \{ m(t), p(t) \}$, which is the creature's size $m(t)$ and its probability of moving $p(t)$. These change as a creature ages to simulate physical growth, according to:
		\begin{align}
		\frac{d m(t)}{dt} &= s\,m(t) \left( 1 - \frac{m(t)}{M} \right),~~~~ m(0) = 0.05M, \\
		p(t)     &= \frac{m(t)}{M} P, 
		\label{eq:mass_evol}
		\end{align}
where we have modeled the movement of the creature as proportional to its mass, corresponding to steps defined by a characteristic length. An exponent in the relation between $p$ and $m$ can be used to represent other geometries or movement characteristics.  

The phenotype defines $\varphi$ and $C$ in the following way. The feeding function $\varphi$ is such that a creature will eat enough from the resource $f$ available at $\mathbf{x}$ to fill its energy storage (of maximum size $m(t)$, the current mass) or, if the resource is not enough for this, whatever is available. The available resource was limited so that a small amount of resource ($0.01K$) always remains in each cell to prevent reaching the (unstable) equilibrium $f = 0$ that is unable to regrow: 
	\begin{equation}
		\varphi(t) = \min \big( \, m(t) - E(t) \, , \, f(\mathbf{x},t) - 0.01K \, \big). 
		\label{eq:phi}
	\end{equation}

The consumption $C$ involves three terms, which represent different bodily needs. Firstly, we have metabolism,  with an allometric dependence on mass that is found in living creatures across many orders of magnitude of body size~\cite{m34}. The second one is kinetic energy, necessary for movement at velocity $v(p)$. We model this as a stochastic process taking the value $1$ when the creature moves from one cell to another (with probability $p$) and $0$ otherwise. Finally, we have the cost of growing, i.e. the power a creature consumes in increasing its mass from birth to adulthood. In mathematical form:
	\begin{equation}
		C(t) = \mu m^{\nicefrac{3}{4}} + \kappa m v^2(p) + s m \left( 1 - \frac{m}{M} \right).
		\label{eq:cons}
	\end{equation}
Coefficients $\mu$ and $\kappa$ in the metabolic and kinetic terms are proportionality coefficients that may be interpreted as efficiencies; they are constant throughout the populations evolution and do not play a relevant role in the model. We have chosen as $\mu = 0.1$ and $\kappa = 0.5$.	

The final mechanisms that need to be defined for the population to undergo natural selection are death and reproduction. Death happens when a creature empties its energy storage ($E=0$) or spontaneously with an age-increasing probability to include the possibility of death by old age. The value of this probability was set to give an overall expected lifetime of $\lambda = 1000$ time steps; we will characterize it thoroughly in Section \ref{sec:res:death}. 

Reproduction proceeds asexually, simulated phenomenologically as follows. First, we require an adult individual ($m(t) > 0.95M$) that keeps its energy above a threshold $T_R = 0.8M$ for $\tau = 100$ time steps. After this initial stage, the body size begins increasing, following a logistic dynamics similar to the one it followed for its own growth, but with a limit value of $2M$. This dynamics halts at $m(t) = 1.05M$, when the offspring separates from the parent, producing two distinct individuals. The brood starts its autonomous existence with mass $0.05M$ and the parent is left with the remaining $M$. The offspring inherits a genome $\bf G'$ which is a mutation of the parent's genome $\bf G$, drawn from a random, uniform distribution centered at $\bf G$ with a maximum mutation of $5\%$ in each gene. We would like to remark that most of the reproduction parameters were chosen arbitrarily and may not represent realistic values. The values were chosen so that the relevant phenomena could be analyzed in a reasonably clear way. The parameters are also left untouched by mutations since it is difficult to establish how their effects might be beneficial in one way but detrimental in others. In absence of optimal values, the system may be susceptible to never reaching a stationary state.

\section{Results}

The hydra model was studied mainly through numerical simulations. No additional assumptions were made upon implementation aside from those described in Section~\ref{sec:age_hyd}. Results were obtained by measuring demographic characteristics like gene distributions and total population, as well as resource availability at periodic intervals during the simulation. The experiments were repeated several times (typically 20) to identify persistent features in the stochastic realizations. The resource recovery speed $r$ was varied over sets of realizations to study the evolution under different conditions. In this section, we present these results and our interpretation of them.

\subsection{Population and genetics}
	\begin{figure}[t]
		\centering
		\includegraphics[width=0.7\textwidth]{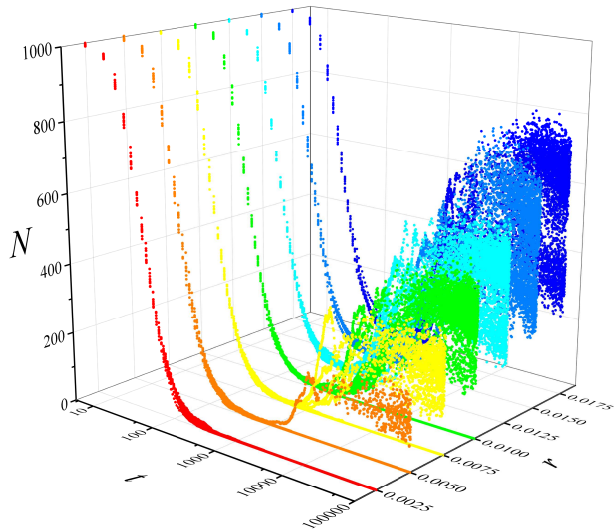}
		\caption{Population dynamics. For each value of the resource recovery $r$, 10 runs of the simulation are displayed, to show the existence of two different stationary states, as evidenced by the formation of two clouds of points. }
		\label{fig:pop_evol}
	\end{figure}

	Figure \ref{fig:pop_evol} shows the temporal evolution of the population for different values of the resource recovery speed, $r$. The simulations begin with $1000$ individuals distributed randomly over both the physical and genetic spaces. The population dynamics shows three distinct regimes (note the logarithmic scale in the time axis). The first one is a wipeout of any genes too unfit for survival, either because of competition with others or because of their own depletion of local resource and energy reservoir. The second one is a population explosion where the surviving genes reproduce while the capacity of the system allows them to do so. The last one is a stationary population value with no genetic change but constant individual renewal. Both transient regimes occur very quickly in terms of elapsed generations, which is reminiscent of the bursts in which evolution is thought to happen in nature\,\cite{raup1986,gould1977,eldredge1988,bak1993}. 

\begin{figure}[t]
		\centering
		\includegraphics[width=0.49\linewidth]{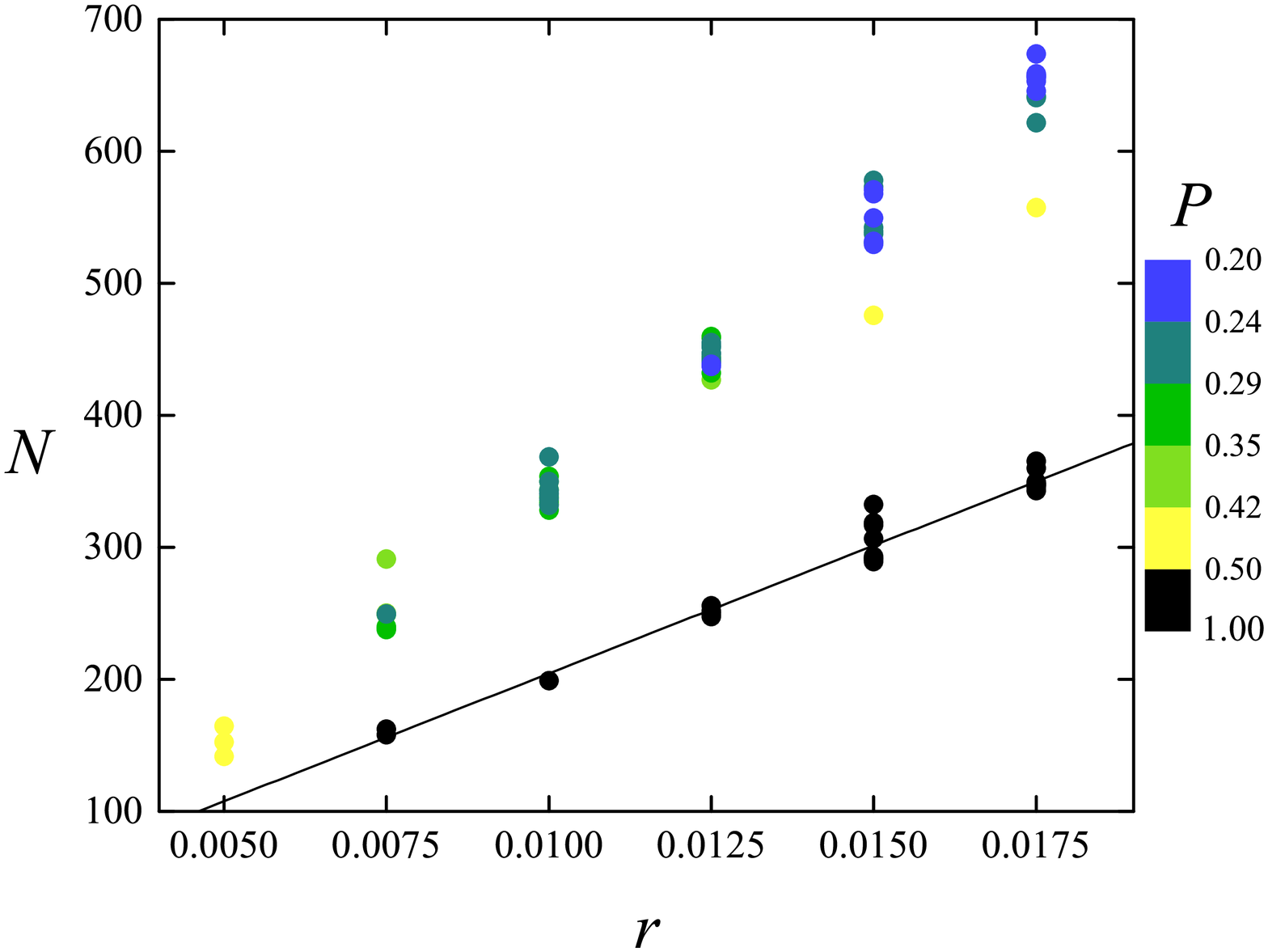}
		\includegraphics[width=0.49\linewidth]{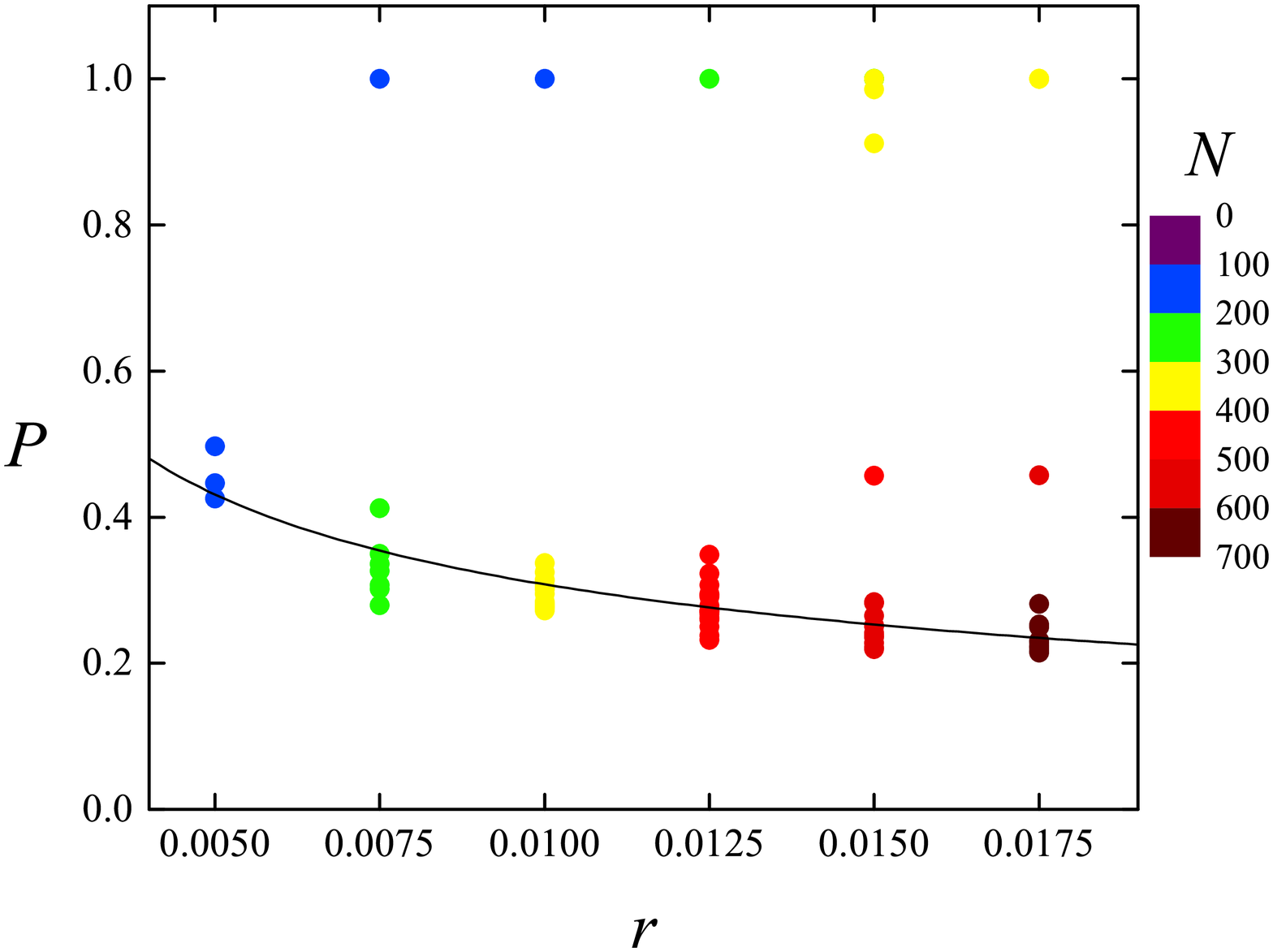}
		\caption{Left: Stationary population vs resource recovery rate. Both branches are linear in $r$. Since the bottom branch is constant in probability of taking a step, its slope can be calculated with the mean-field equation~(\ref{eq:norm_av_res}) obtained in Section~\ref{sec:res:mean_field}. Right: Stationary adult speed vs resource recovery rate. The top branch is constant in $r$, $P=1$ and corresponds to the bottom branch of the left figure. We found that the bottom branch obeys $P \propto r^{-1/2}$ and corresponds to the top branch of the left figure.}
		\label{fig:pop_spd_rcv}
	\end{figure}

	From now on, let us focus on the stationary state unless stated otherwise. A faster-growing resource means that there is more food available and less need to move, so the population increases and the adult speed decreases as the resources recovery speed does. Additionally, both present two branches that can be interpreted as two possible traits or species that may emerge given a recovery speed. These features of the stationary regime are shown in Fig.~\ref{fig:pop_spd_rcv}. Bear in mind that the top branch of one plot corresponds to the bottom one of the other. This means that a larger population must consume less energy for the resource to be able to sustain it. We will address this inverse relationship more in detail in Section \ref{sec:res:mean_field}.

	The adult mass, on the other hand, reaches a stationary value that is independent of the recovery rate of the resource, within the precision achieved in the simulations. In a phenomenological linear fit $M = M_o + ar$, we obtained $M_o = 0.0717\pm0.0004$ and $a = 0.02\pm0.03$. For values of $r$ in the simulated range, the error in the constant term is of the same order as the value of the linear one, making the latter negligible. Meanwhile, the growth speed $s$ increased monotonously in all the simulations. The specific form in which this parameter changed was the same for all simulations regardless of the value of $r$, meaning that a faster growth speed always increases chances of survival. The effect of this gene seems to be purely in the amount of time over which a fixed amount of energy, determined by the mass gene, is expended on growth. Since the behavior of both of these parameters is untouched by the parameters we varied, we have mostly ignored them in this analysis.  

\begin{figure}
		\centering
		\includegraphics[width=0.6\linewidth]{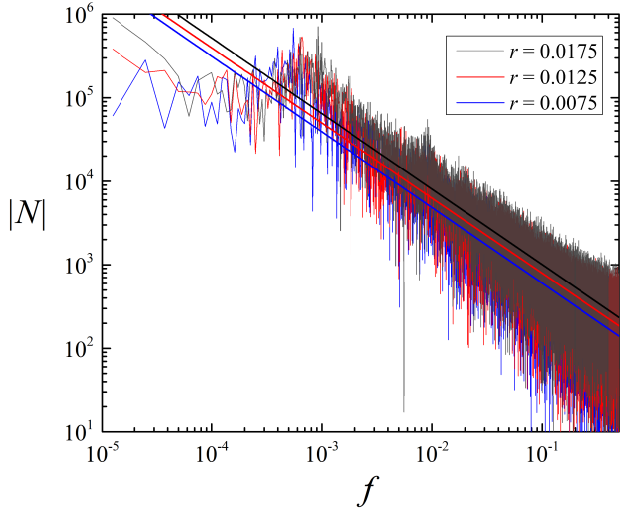}
		\caption{Power spectrum of population fluctuations obtained from population dynamics in the stationary state.}
		\label{fig:pow_spec_N}
	\end{figure}

	Figure \ref{fig:pow_spec_N} shows the power spectrum obtained from the population dynamics. It displays a power-law behavior of the type $f^{-\alpha}$ with $\alpha \approx 1$. This means that population fluctuations are very similar to the ones obtained in simpler models~\cite{bak1993} in which fitness has no physical interpretation but natural selection takes place nevertheless.

\subsection{Altruistic behavior}

\begin{figure}[t]
		\centering
		\includegraphics[width=0.6\linewidth]{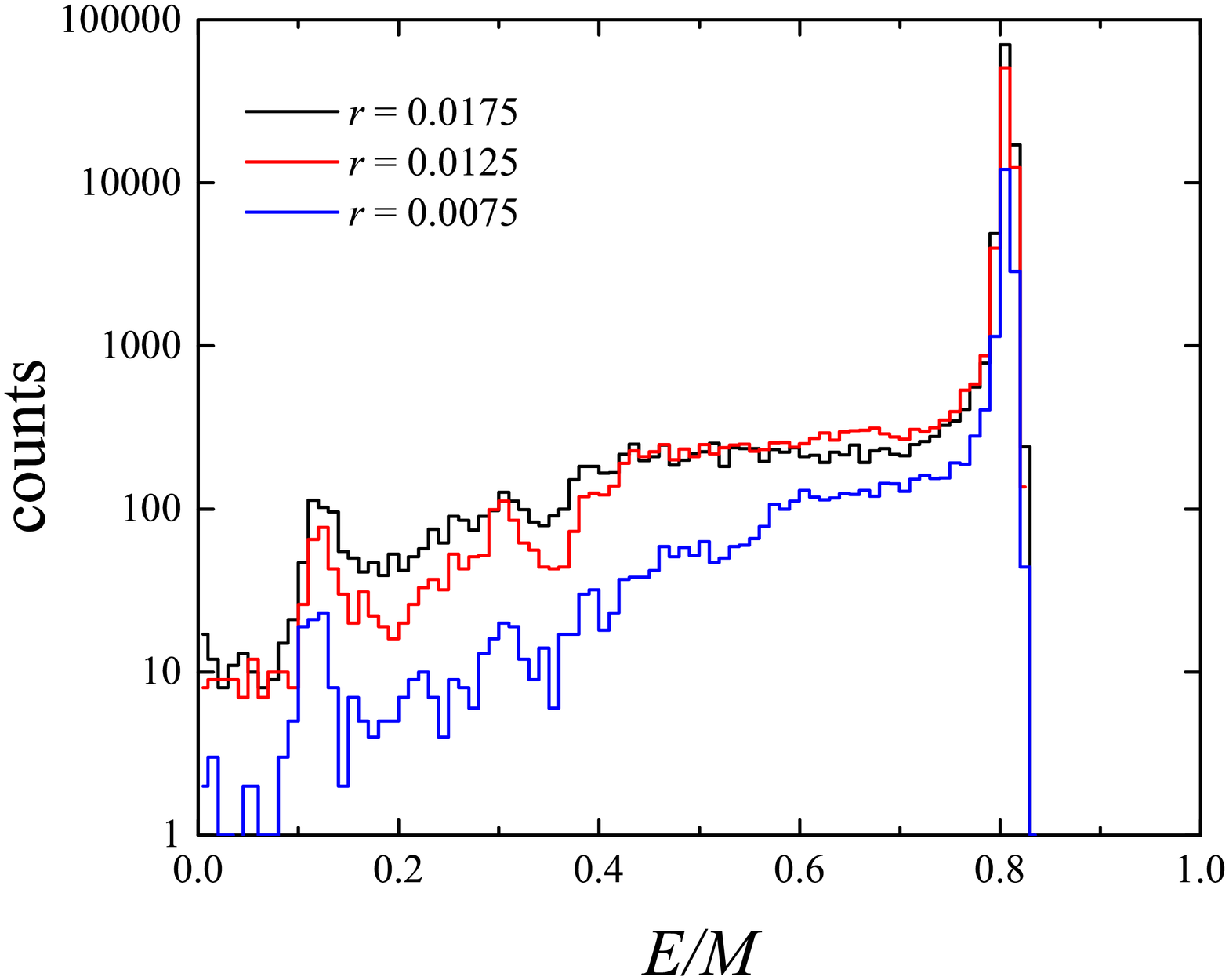}
		\caption{Energy distribution measured over a stationary population. Note that most individuals have energy a little over the reproduction threshold $T_R/M = 0.8$. Three values of the recovery are shown, as indicated in the legend.}
		\label{fig:engy_dist}
	\end{figure}

	We have observed the emergence of a rudimentary altruistic behavior from egoistic rules, in the sense proposed by the ``selfish gene'' theory~\cite{williams1966,dawkins1976}. The feeding function modeled by Eq.~(\ref{eq:phi}) might suggest that feeding would always be as large as possible, limited only by the capacity of the creature or the availability of the resource. However, interactions in the system manage to provide a setting in which individuals can maintain the minimum amount of energy needed to reproduce instead of the maximum energy possible. Figure~\ref{fig:engy_dist} portrays this effect by giving an estimation of the energy distribution throughout the population. Instead of being concentrated around $E/M = 1$, it stays just above the reproduction threshold $T_R/M = 0.8$ (note the logarithmic scale). This is altruistic both towards other creatures, that are left with more available resource, and with the resource, by preserving part of it for its recovery. 

\subsection{Age distribution and death}
\label{sec:res:death}

	We will now fulfill our promise to characterize spontaneous death more thoroughly. First, consider the probability of being dead at age $t$, which we will denote $P_D(t)$. It is the cumulative distribution of the probability of death at age $t$, $p_D(t) = dP_D(t)/dt$. The probability of death is also related to the rate of lethal events, $\rho_D(t)$. The latter represents dangers that may or may not kill an individual at each moment in time. Simply put, the rate of lethal events is to the probability of death as the outcome of a single coin toss is to the process of tossing a set of coins until they fall heads up. The probability of being dead is the proportion of coins that have stopped being tossed. Since it is a time-dependent Poisson process, we have:
		\begin{align}
			p_D(t)  =& \rho_D(t) \, e^{-\int_0^t  \rho_D(t') dt' }, \\
					=& -\frac{d}{dt} \left( e^{-\int_0^t  \rho_D(t') dt'}\right),
		\label{eq:death_dist1}
		\end{align}
and:
		\begin{align}
			\ln \left( 1 - P_D(t) \right) =& \int_0^t  \rho_D(t') dt', \label{eq:death_dist1.5}\\
			\Rightarrow \rho_D(t) =& \frac{p_D(t)}{ 1 - P_D(t) },
		\label{eq:death_dist2}
		\end{align}
where in the left hand side of Eq.~(\ref{eq:death_dist1.5}) we made use of the initial condition $P_D(0)=0$.

	As mentioned before, we assume that our creatures have a probability of dying spontaneously due to aging. This occurs in such a way as to provide a lifetime of $\lambda = 1000$ time steps, and a slowly increasing $P_D(t)$. We chose a logistic sigmoid function for this purpose:
		\begin{align}
		P^\text{age}_D(t) = \frac{1}{1 + e^{-2(t - \lambda)/\lambda}}, \\
		p^\text{age}_D(t) = \frac{2}{\lambda}\frac{e^{-2(t - \lambda)/\lambda}}{ \left( 1 + e^{-2(t - \lambda)/\lambda} \right)^2}, \\
		\rho^\text{age}_D(t) = \frac{2}{\lambda}\frac{1}{ 1 + e^{-2(t - \lambda)/\lambda}}. 
		\label{eq:prob_death}
		\end{align}

\begin{figure}[t]
		\centering
		\includegraphics[width=0.6\linewidth]{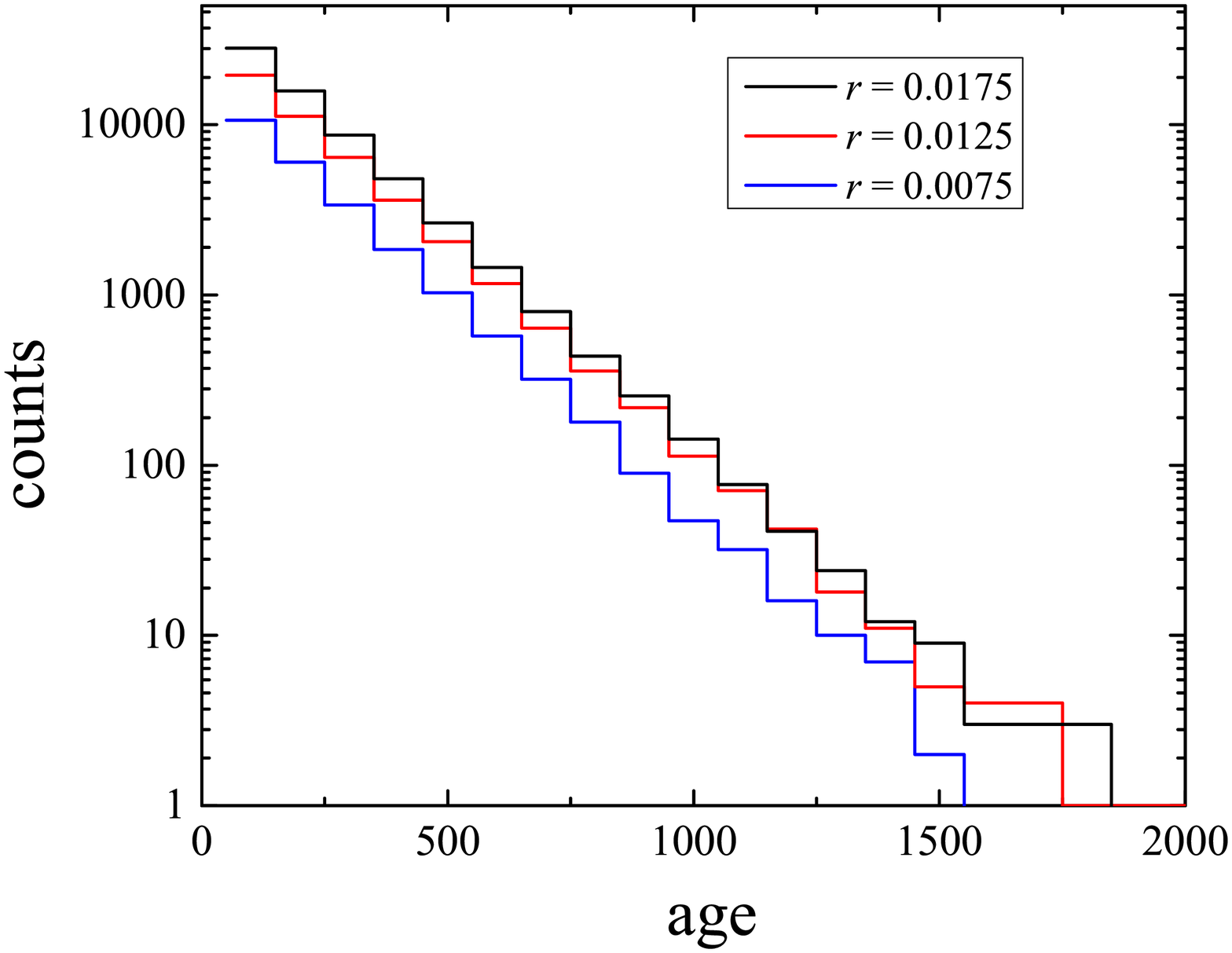}
		\caption{Age distributions normalized to the total population in the stationary regime of the system's evolution. Since they are almost perfect exponentials, the mortality rate given by using Eq.~(\ref{eq:age_dist}) is $\rho_D=0.055\pm0.003$. Its inverse is the expected age, $\tau = 181\pm9$.}
		\label{fig:age_dist}
	\end{figure}

	Since individuals are also susceptible to dying from lack of energy, the total rate of deadly events is $\rho_D(t) = \rho_D^{\text{age}}(t) + \rho_D^{E}(t)$. We can relate it to the age distribution in the population, which we will denote by $z(t)$, where $z(t)dt$ is the amount of creatures of age $t$. If $N$ is the total population, $\int z(t)dt = N$. The age distribution evolves according to:
		\begin{align}
		z(t+dt) &= z(t) - z(t) \rho_D(t) dt, \\
		d_t z(t) &= - z(t) \rho_D(t), \\
		\Rightarrow \rho_D(t) &= - d_t \ln \left( z(t) \right).
		\label{eq:age_dist}
		\end{align}
Equation (\ref{eq:age_dist}) expresses the total rate of deadly events in terms of age distribution. Age distributions are shown in Fig.~\ref{fig:age_dist} in the steady-state of evolution for three values of the resource recovery parameter $r$. Since they are almost perfect exponentials, equation (\ref{eq:age_dist}) implies that the death rate is constant. Its inverse is the average overall lifetime of an individual in the population. For all three cases, the obtained lifetime is $\tau = 181 \pm 9$. We can therefore conclude that death is governed by lack of energy, not aging. Lifetime is enough for growth and reproduction, but well under the $\lambda = 1000$ available if only death by old age is considered. Again, the system has produced an altruistic behavior of sorts arising from selfish rules.

\begin{figure}[t]
		\centering
		\includegraphics[width=0.7\linewidth]{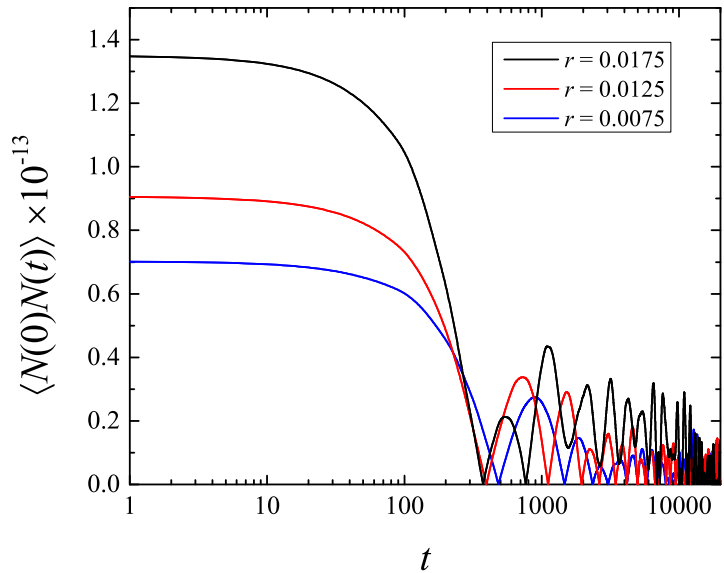}
		\caption{Population autocorrelation obtained from the population power spectrum. Given the different population sizes for different values of $r$, the autocorrelation is measured on deviations from the population average.}
		\label{fig:autocorr_N}
	\end{figure}

An interesting way to view individual lifetime is by looking at the autocorrelation of the population. It can be obtained by applying the Wiener-Khinchin theorem~\cite{vankampen} to the power spectrum shown in Fig.~\ref{fig:pow_spec_N}. The result is shown in Fig.~\ref{fig:autocorr_N}. It is interesting to note how the correlation drops quickly just above $t = 100$, which we attribute to the fact that almost all creatures in the simulation die in this range, and therefore the system loses coherence throughout the creatures.

\section{Mean-field dynamics}
\label{sec:res:mean_field}

	In this section we present a mean-field description of the resource dynamics, validated with the results obtained in simulations.

	The average consumption rate of a creature $c$ with genotype $\mathbf{G} = \{ M, P, s \}$ and phenotype $\mathbf{F} = \{ m, p \}$ is $\langle C \rangle = C(\mathbf{F}_c) = \mu m^{\nicefrac{3}{4}} + \kappa m p + s m \left( 1 - \nicefrac{m}{M} \right)$. For a population of creatures $\mathcal{P}: c \in \mathcal{P}$ of which a subset $\mathcal{P}(\mathbf{x})$ are at a site $\mathbf{x}$ at a certain time, their combined average consumption during a short interval $\Delta t$, $a(\mathbf{x})$, is: 
	\begin{equation}
			a(\mathbf{x}) = \min \left( \sum_{\mathcal{P}(\mathbf{x})} C(\mathbf{ F}_c)\Delta t , f(\mathbf{x}) - 0.01K \right).
		\label{eq:tot_cons}
	\end{equation}
Note that $\lim_{\Delta t \rightarrow 0} a(\mathbf{x}) = \sum_{\mathcal{P}(\mathbf{x})} C(\mathbf{F}_c) dt$ unless the resource is strictly $0.01K$. Thus, in the continuous time limit, the average resource dynamics can be obtained as follows:
	\begin{align}
		\frac{d \langle f \rangle_{\mathbf{x}}}{dt} &= \frac{1}{L^2} \sum_{\mathbf{x}}\left[ r f(\mathbf{x}) \left( 1 - \frac{f(\mathbf{x})}{K} \right) - \frac{a(\mathbf{x})}{dt}\right], \notag\\
		&= \frac{1}{L^2} \sum_{\mathbf{x}} r f(\mathbf{x}) \left( 1 - \frac{f(\mathbf{x})}{K} \right) - \frac{N}{L^2} \frac{1}{N} \sum_{\mathbf{x}} \sum_{\mathcal{P}(\mathbf{x})} C({\bf F_c}), \notag\\
		& = r \left[ \langle f \rangle_{\mathbf{x}} \left( 1 - \frac{\langle f \rangle_{\mathbf{x}}}{K} \right) - \frac{\text{var}(f)}{K} \right] - \frac{N}{L^2} \langle C \rangle_{\mathcal{P}}.
		\label{eq:av_res}
	\end{align}	
In Eq.~(\ref{eq:av_res}) we have assumed that $r$ and $K$ are homogeneous throughout the landscape for simplicity, but of course this does not need to be the case, and the averages should be taken accordingly. 
If we normalize the resource by its local carrying capacity $K$, defining $n=f/K$, the dynamics becomes:
	\begin{equation}
		\frac{d \langle n \rangle_{\mathbf{x}}}{dt} = r \left[ \langle n \rangle_{\mathbf{x}} \left( 1 - \langle n \rangle_{\mathbf{x}} \right) - \text{var}(n) \right] - \frac{N}{L^2} \frac{\langle C \rangle_{\mathcal{P}}}{K}.
		\label{eq:norm_av_res}
	\end{equation}
This equation resembles the prey's dynamics in a predator-prey system, where the normalized resource takes the role of the prey, and $N/L^2$ is the normalized predator population. The interaction is a consumption of prey at a rate given by the last term in Eq.~(\ref{eq:norm_av_res}), represented by $C$, since this comes from the consumption rate $a$, which is explicitly related to $f$, the ``actual'' prey.

\begin{figure}[t]
		\centering
		\includegraphics[width=0.7\linewidth]{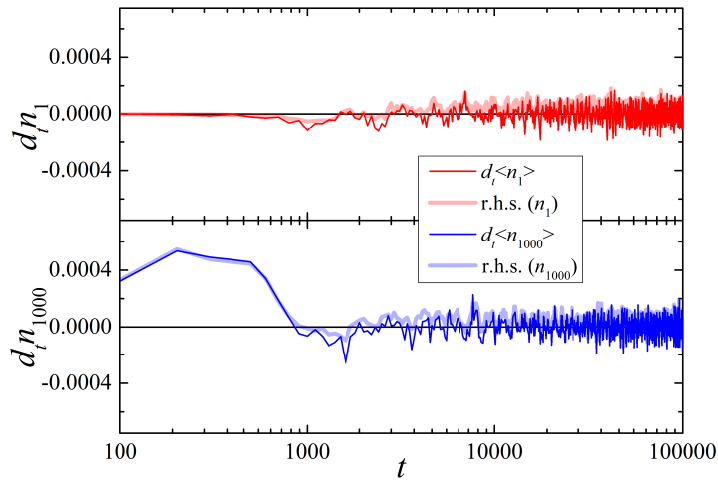}
		\caption{Resource dynamics for different initial conditions. Both sides of Eq.~(\ref{eq:norm_av_res}) are shown. Top: initial condition is a single individual. Bottom: initial condition has 1000 individuals. }
		\label{fig:res_dyn}
	\end{figure}

	Figure~\ref{fig:res_dyn} shows both sides of Eq.~(\ref{eq:norm_av_res}) as measured from numerical simulations of the system. Two simulations are shown, identical in all except the initial population. The first one, denoted by $n_1$, had an initial population of only $1$ individual. We observe that it starts and stays in equilibrium ($d_t \langle n \rangle_{{\bf x}} = 0$). The second one, denoted by $n_{1000}$ has an initial population of $1000$ and achieves equilibrium after a transient of $t\approx 1000$ time steps. After that, all curves follow similar paths. In both cases the measured left and right-hand sides of Eq. (\ref{eq:norm_av_res}) coincide almost perfectly, supporting the validity of the mean-field dynamics. The initial difference between the $n_{1000}$ curves are a product of different sampling rate when measuring $d_t \langle n \rangle_{{\bf x}}$ and the corresponding right-hand side.

	To complete the description of the dynamics and calculate the system's equilibria we need an equation for $d_t N$. 
The dynamics of $N$ is given by the balance of births and deaths in the population. Since both of these are controlled by the energy of the creatures, which fluctuates because of the resource, births and deaths are stochastic processes. For this reason we must first advance the understanding of the resource dynamics and its fluctuations around its mean value.

	\subsection{Resource distribution}
	\label{sec:res_dist}

	\begin{figure}[t]
		\centering
		\includegraphics[width=0.48\linewidth]{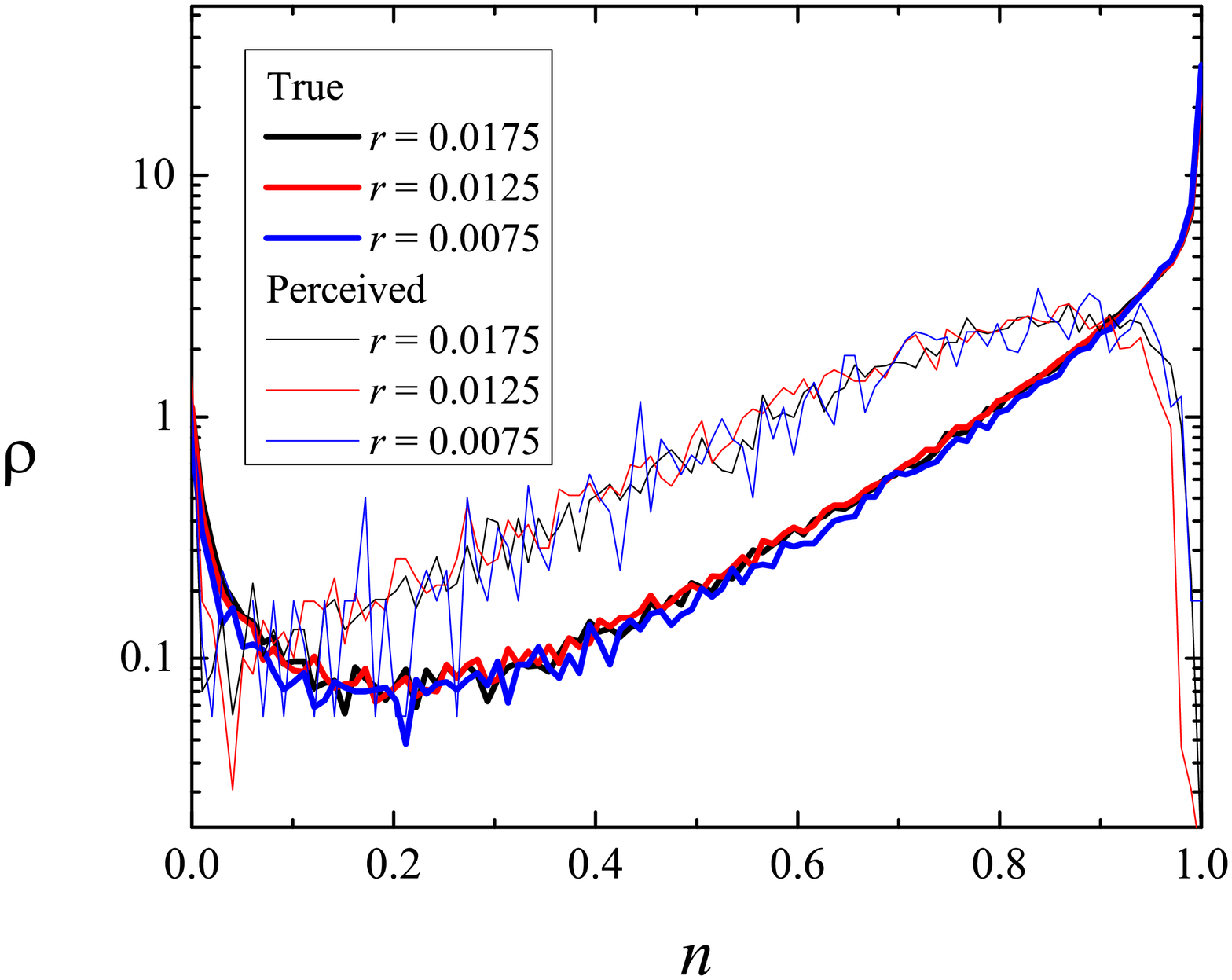}
		\includegraphics[width=0.50\linewidth]{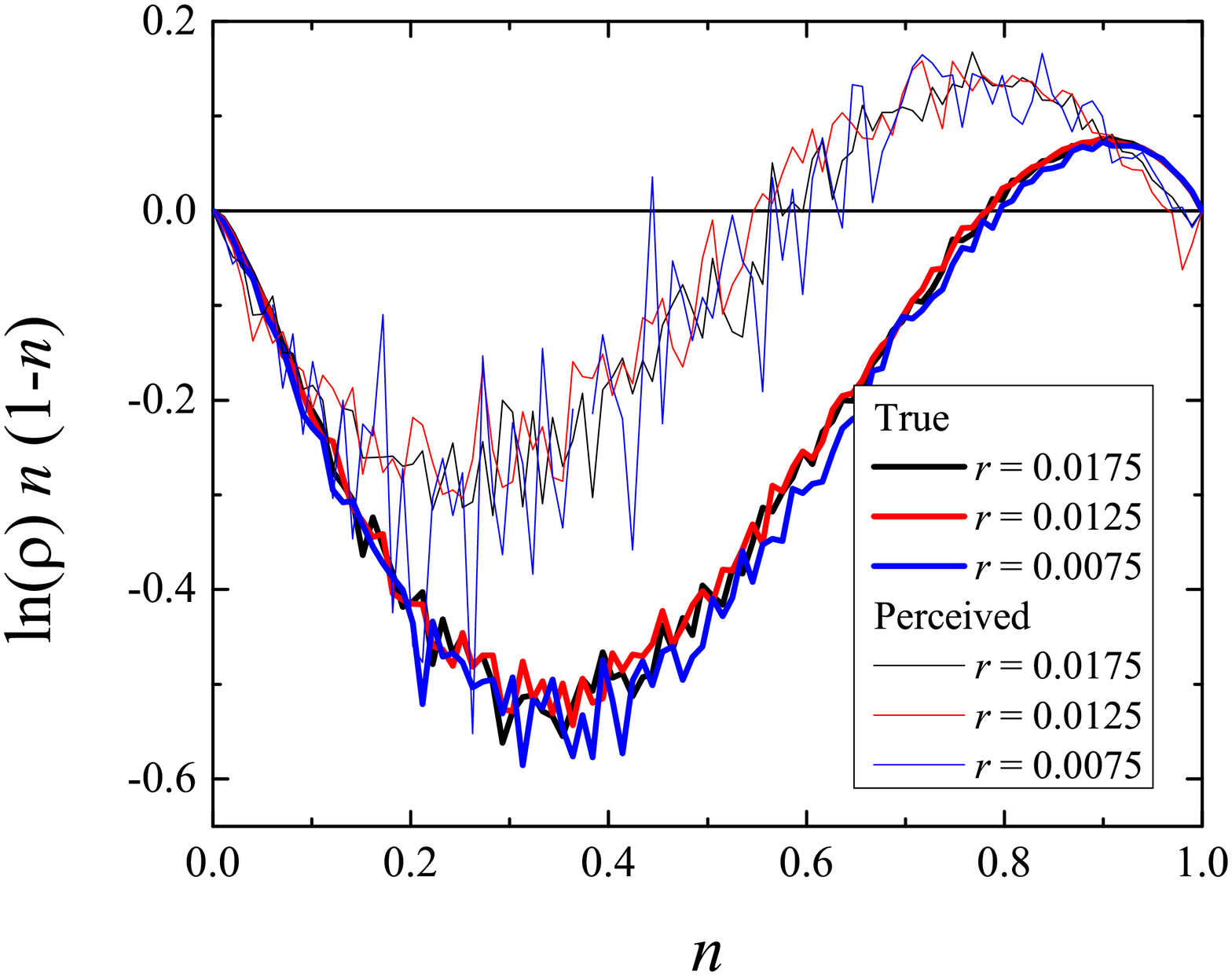}
		\caption{True and perceived resource distributions in the stationary state for several values of recovery rate. Right: transformed distributions, to remove divergences.}
		\label{fig:res_dist}
	\end{figure}

	Figure \ref{fig:res_dist} (left, thick lines) shows the resource distribution in equilibrium for different values of $r$. Since the distributions have sharp spikes at both extremes of the resource range, $0$ and $1$, we prefer to transform them by taking the logarithm and multiplying by $n(1-n)$ to eliminate any possible divergences. The right panel (thick lines) shows these transformed resource distributions.  

	Since creatures need to be in a cell to consume its resource, the distribution they perceive is the conditional distribution $\rho(n|\text{o})$ of resource, given that the cell is occupied. Figure \ref{fig:res_dist} (left, thin lines) shows the corresponding perceived distributions as measured from the simulations (for the same values of $r$). Although their concavities are different, as well as their divergences, the perceived distributions display a shape similar to the true ones when transformed accordingly, as shown in Fig.~\ref{fig:res_dist} (right, thin lines).

\begin{figure}[t]
		\centering
		\includegraphics[width=0.6\linewidth]{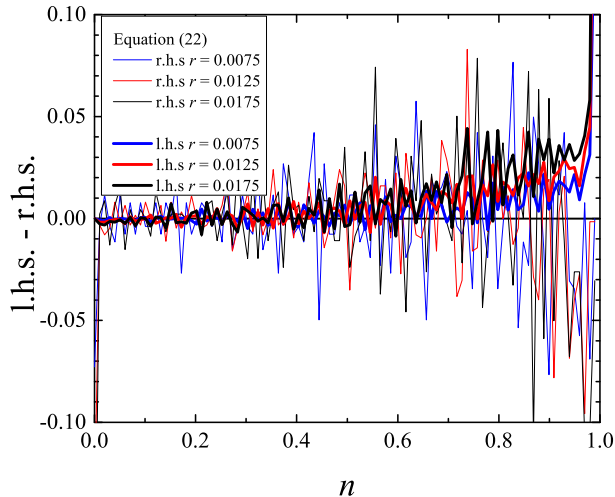}
		\caption{Dynamics of the resource, showing both sides of Eq.~(\ref{eq:res_dist_eq}). For values of $n \approx 1$, the curves separate, indicating that the system is not in equilibrium in this region.}
		\label{fig:biased_res_dist}
	\end{figure}

The true and perceived distributions are related by consumption and recovery, and in equilibrium we can derive their relation explicitly. First we decompose the temporal evolution of $\rho$ according to its dependence on the resource $n$: 
\begin{equation}
	\frac{d\rho(n)}{dt} = \frac{d\rho(n)}{dn} \frac{dn}{dt}.
\end{equation}
Conditional probabilities allow us to write this in terms of occupied (o) and not occupied ($\neg$o) sites, and equal to zero in equilibrium:
\begin{equation}
	\frac{d\rho(n)}{dt}	= \frac{d\rho(n | \text{o})}{dn} \rho(\text{o}) \left.\frac{dn}{dt}\right|_{\text{o}} + \frac{d\rho(n | \neg \text{o})}{dn} \rho(\neg \text{o}) \left.\frac{dn}{dt}\right|_{\neg \text{o}} = 0.
\label{eq:drhodt}	
\end{equation}
Now, the dynamics of unoccupied sites is just the logistic recovery, $d_t n|_{\neg o} = rn(1 - n)$, while that of occupied sites is a balance of recovery and consumption, $d_t n|_{o} = d_t n|_{\neg o} - \langle C \rangle_{\mathcal{P}}/K$. Using these in Eq.~(\ref{eq:drhodt}), together with the fact that $\rho(o)\rho(n|o) + \rho(\neg o)\rho(n|\neg o) = \rho(n)$, elementary algebra allows to derive:
\begin{equation}
	rn(1-n)\frac{d\rho(n)}{dn} = \frac{\langle C \rangle_{\mathcal{P}}}{K} \frac{N}{L^2} \frac{d\rho(n | o)}{dn}. 
\label{eq:res_dist_eq}
\end{equation}
Both sides of Eq.~(\ref{eq:res_dist_eq}) are shown in Fig.~\ref{fig:biased_res_dist}, calculated from measurements performed on simulations for the three values of $r$ shown in previous figures. We can see that both bundles of curves coincide, within fluctuations, for a very wide range of values of $n$, separating when $n$ approaches unity. It is worth noting that Eq.~(\ref{eq:res_dist_eq}) fails as $n \to 1$, which indicates that this region never reaches equilibrium fully.  

\section{Discussion and conclusions}

    We have proposed and analyzed an agent-based model in which the energy available to living creatures determines the processes that characterize life: birth, reproduction and death. The dynamics of energy itself is defined by genes and indirect competition with other creatures through a common resource. Genes determine body size (i.e., mass and energy reservoir size), movement speed across the domain of the resource and growth speed from birth to adulthood. Small, random mutations upon asexual reproduction allow different genetic combinations to compete with one another and therefore natural selection emerges with only a shared resource and random mutations.

    It is interesting to relate an intuitive notion of the intensity of natural selection with the system's capacity to sustain a population. In this sense, natural selection is the force that eliminates individuals while the recovery speed of the resource keeps them alive. As opposed to some artificial selection models in which a quality of individuals is explicitly given and so the best can always be picked, natural selection doesn't guarantee a surviving population. This is seen for simulations starting with a large genetic diversity and small values of resource recovery speed. For larger recovery speeds, an initial mass extinction is observed, followed by an explosion of the successful genes. Even when the resource recovers instantly (i.e., the resource is limitless), some genes are still unfit to survive and the initial extinction is still observed. We conclude that natural selection decreases with the capacity of a system but is always present to some degree.

\begin{figure}[h]
        \centering
        \includegraphics[width=0.7\linewidth]{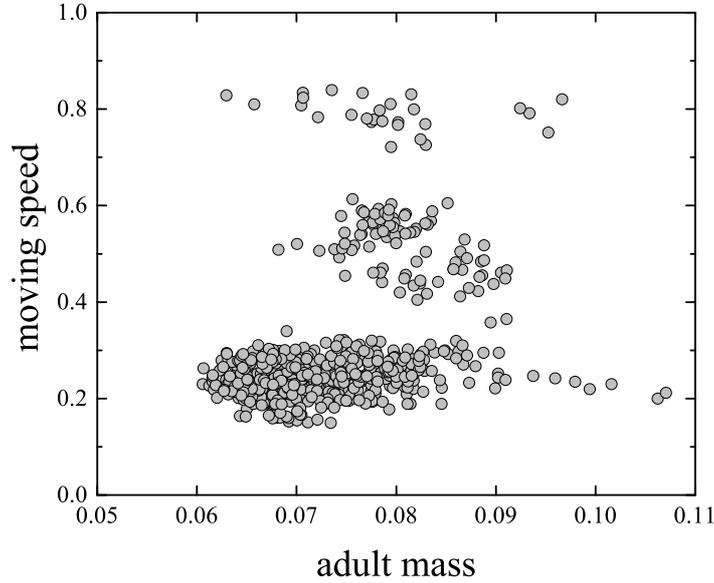}
        \caption{Brief coexistence of several distinct species in the initial, transient regime of a typical simulation. The recovery speed is $r = 0.0175$ and the state occurs about halfway through the transient period.}
        \label{fig:gene_coex}
    \end{figure}

    After the initial extinction and population explosion, we found that two distinct genetic groups emerge in the stable state of the system. They are separated by their moving speed but indistinguishable in their mass. The growth speed, on the other hand, turned out to be a gene with which fitness increased monotonously, so it never reached a stationary value but didn't affect the system. Since a faster movement implies more energy consumption, fast populations were found to be smaller. However, only one of these ``species'' survived in each realization. Species were observed to compete briefly in the initial transient period, as shown in Fig.~\ref{fig:gene_coex}, but one of the groups always eliminated the others in accordance with the competitive exclusion principle~\cite{murray} and punctuated equilibrium.

    Simulations starting with a single individual stayed in equilibrium throughout the evolution of the system, even during population growth. They also exhibited stationary states at resource recovery speeds slower than ones at which larger populations completely died out. Additionally, these populations always converged to low energy consumption populations. This suggests that the high energy consumption species found earlier are a result of a non-equilibrium situation, and thus the system is very sensitive to initial conditions. Another noteworthy conclusion is that, according to this model, life beginning from a single organism seems to be more likely than life beginning from a larger group, even at the cost of genetic diversity.

    The final result we would like to remark from simulations is that, even though the rules followed by each individual were set to be selfish, the system managed to evolve into a somewhat altruistic setting. This was observed in two instances. The first regards the amount of resource consumed, where creatures were supposed to eat enough to fill their energy reservoirs but ended up keeping them at the minimum necessary to reproduce. The second relates to lifetime, that could end by old age after around ten reproductions, but ended up limited to the time necessary for just one reproduction. This makes perfect sense because the population must remain constant, but it also shows that the probability of producing offspring, usually associated with fitness, is not maximized under natural selection with a shared resource. The population size, used as a fitness measure in works such as~\cite{menczer1996a,menczer1996b} does not seem to be maximized either, since for the same conditions we have found realizations that attain a stationary state with a large population and low consumption, and others with small populations and high consumption. It is worth emphasizing that this single-offspring situation was not programmed in the model, but emerged from constraints and interactions.

    Although the model is somewhat artificial (agents are, quite literally, spherical cows), our study demonstrates that all of the phenomena mentioned above are actually parts of a whole. They can be obtained by considering that the genetic code of an individual determines how energy is obtained and employed to harvest more. An open question is whether more detailed models of genetic dependence produce more realistic results in terms of surviving genes. Another is if predator-prey systems are equivalent to resource-population systems where the resource is controlled by genes in the same way as the population.

    We have taken an initial step towards answering the first question by adding a rudimentary intelligent behavior to the population. With it, movement speed depends linearly on the amount of energy in a creature's reservoir instead of being constant. This allows creatures to move more when they are hungry and less when they are full and vice-versa. The first case is more energy efficient but dangerous when there is little food nearby, while the second is less efficient but could be safer in terms of keeping energy levels high. Both cases have their advantages and drawbacks and therefore we cannot predict  which will appear. However, it is known that the flatworm \textit{Dugesia tigrina} exhibits this rudimentary intelligence and behaves as in the first case~\cite{DugesiaTigrina}. In our simulations, we found that the creatures in the stationary state were those with genes that made them move more when they were hungry and less when they were not. Since this agrees with the real-life scenario we conclude that, as genetic models get better, the simulation results will also get more realistic.

    Finally, we would like to mention the consequences of the mean-field dynamics analysis. Much can be read from Eq.~(\ref{eq:norm_av_res}), describing the dynamics of the mean normalized resource, which we copy here:
	\begin{equation}
		\frac{d \langle n \rangle_{\mathbf{x}}}{dt} = r \left[ \langle n \rangle_{\mathbf{x}} \left( 1 - \langle n \rangle_{\mathbf{x}} \right) - \text{var}(n) \right] - \frac{N}{L^2} \frac{\langle C \rangle_{\mathcal{P}}}{K}.
		\label{eq:mean_field_conc}
	\end{equation}

    First, Eq.~(\ref{eq:mean_field_conc}) resembles the dynamics of a prey, $\langle n \rangle$, faced with a population of $N$ predators consuming them at a rate $\langle C \rangle / K L^2$, as mentioned above. This interpretation suggests that our model is  similar to a predator-prey system. It also makes altruistic behavior (keeping energy at the minimum necessary) beneficial not only for the rest of the population (other predators), but also for the resource (the ``prey''). Of course, this is a consequence of the population needing the resource (or predators needing prey) to survive. However, since we observed populations with higher energy consumption, we can conclude that this form of altruism isn't necessary, merely better.

    Second, note that the growth term of the resource in Eq.~(\ref{eq:mean_field_conc}) is reduced by the variance of the resource. This means that smaller fluctuations in the resource distribution at the cost of a smaller expected value, sometimes will be more beneficial for the population. This effect is definitely non-negligible since an estimation of the population that does not include this term, $N \approx rf(1-f)L^2/C \approx 500$ ($r = 0.01$, $P = 0.33$, $f = 0.85$), results in an error of around $70\%$ compared to the observed population of $N \approx 300$.

    Finally, since all real-world food sources have a finite size, their growth is limited. Therefore a logistic equation, as a second-order expansion, is the simplest realistic assumption we can make. The variance term in Eq.~(\ref{eq:mean_field_conc}) stems from the nonlinear term in this expansion and therefore all real-world resources will suffer from it. Moreover, resource distribution is shaped by consumption, so the efficiency of a species can be characterized through that variance. In this sense, more ``intelligent'' (better distributed) consumption will lead to faster resource recovery and therefore a larger population. This may explain why the first great revolution human beings experienced, the invention of agriculture, was so important~\cite{eshed2004, barker2009}. Since we could suddenly control the dynamics of our source of energy, we could minimize variance much more effectively, ensuring that everyone has enough food and allowing the population to grow. 

    We have set out trying to describe a model that produces natural selection from first principles, namely energy conservation and its flow through living beings. In the process, we have discovered many other known results and characteristics associated with the theory of evolution. We consider this a successful feat, but there is still much work to be done. It is not clear how exactly these phenomena emerge from the underlying process, nor how they are influenced when specifics are changed. We hope to address these questions and any new ones that may appear in future work. 
    
    \section*{Acknowledgments}

We recognize partial support through the following grants: CONICET PIP 112-2017-0100008 CO, UNCUYO SIIP 06/C546, ANPCyT PICT-2018-01181. We also thank S. Bouzat for fruitful discussions.
	
\bibliographystyle{unsrt}
\bibliography{ref}

\end{document}